\documentclass[twoside,twocolumn,9pt]{article}
\usepackage{extsizes}
\usepackage[super,sort&compress,comma]{natbib} 
\usepackage[version=3]{mhchem}
\usepackage[left=1.5cm, right=1.5cm, top=1.785cm, bottom=2.0cm]{geometry}
\usepackage{balance}
\usepackage{mathptmx}
\usepackage{sectsty}
\usepackage{graphicx} 
\usepackage{lastpage}
\usepackage[format=plain,justification=justified,singlelinecheck=false,font={stretch=1.125,small,sf},labelfont=bf,labelsep=space]{caption}
\usepackage{float}
\usepackage{fancyhdr}
\usepackage{fnpos}
\usepackage[english]{babel}
\addto{\captionsenglish}{%
  
}
\usepackage{array}
\usepackage{droidsans}
\usepackage{charter}
\usepackage[T1]{fontenc}
\usepackage[usenames,dvipsnames]{xcolor}
\usepackage{setspace}
\usepackage[compact]{titlesec}
\usepackage{hyperref}
\usepackage{adjustbox}
\usepackage{amsmath,amssymb}
\usepackage{epstopdf}

\definecolor{cream}{RGB}{222,217,201}

\newcommand{\E}[1]{\langle #1\rangle}
\newcommand{\Es}[1]{\langle #1\rangle_{\rm s}}

\newcommand{\rmd}{\mathrm{d}}

\newcommand{\abs}[1]{\left\lvert #1 \right\rvert}

\newcommand{\nus}{\nu_{\rm s}}

\newcommand{\ps}{p_{\rm s}}
\newcommand{\js}{j_{\rm s}}

\newcommand{\var}{{\rm var}}

\usepackage{color}\usepackage{xcolor}
\definecolor{C0}{rgb}{0.12156862745098039, 0.4666666666666667, 0.7058823529411765}
\definecolor{C1}{rgb}{1.0, 0.4980392156862745, 0.054901960784313725}
\definecolor{C2}{rgb}{0.17254901960784313, 0.6274509803921569, 0.17254901960784313}
\definecolor{C3}{rgb}{0.8392156862745098, 0.15294117647058825, 0.1568627450980392}
\definecolor{C4}{rgb}{0.5803921568627451, 0.403921568627451, 0.7411764705882353}
\definecolor{C5}{rgb}{0.5490196078431373, 0.33725490196078434, 0.29411764705882354}
\definecolor{inferno0}{rgb}{0.087411, 0.044556, 0.224813}
\definecolor{inferno1}{rgb}{0.379001, 0.076253, 0.432719}
\definecolor{inferno2}{rgb}{0.658463, 0.178962, 0.372748}
\definecolor{inferno3}{rgb}{0.894305, 0.353399, 0.193584}
\definecolor{inferno4}{rgb}{0.987622, 0.64532 , 0.039886}

\colorlet{mylinkcolor}{blue!66!black!80}
\hypersetup{colorlinks = true, linkcolor = {mylinkcolor}, linkcolor = {mylinkcolor}, citecolor = {mylinkcolor}, urlcolor = {mylinkcolor}}
\definecolor{grey}{rgb}{0.6,0.6,.6}
\definecolor{darkgrey}{rgb}{0.4,0.4,.4}
\definecolor{darkgreen}{rgb}{0,0.4,0}
\definecolor{lightgreen}{rgb}{0,0.7,0}
\definecolor{darkred}{rgb}{0.5,0,0}
\newcommand{\blue}[1]{{\color{black}#1}}
\newcommand{\red}[1]{\blue{#1}}

\begin{document}

\pagestyle{fancy}
\thispagestyle{plain}
\fancypagestyle{plain}{
\renewcommand{\headrulewidth}{0pt}
}

\makeFNbottom
\makeatletter
\renewcommand\LARGE{\@setfontsize\LARGE{15pt}{17}}
\renewcommand\Large{\@setfontsize\Large{12pt}{14}}
\renewcommand\large{\@setfontsize\large{10pt}{12}}
\renewcommand\footnotesize{\@setfontsize\footnotesize{7pt}{10}}
\makeatother

\renewcommand{\thefootnote}{\fnsymbol{footnote}}
\renewcommand\footnoterule{\vspace*{1pt}%
\color{cream}\hrule width 3.5in height 0.4pt \color{black}\vspace*{5pt}} 
\setcounter{secnumdepth}{5}

\makeatletter 
\renewcommand\@biblabel[1]{#1}            
\renewcommand\@makefntext[1]%
{\noindent\makebox[0pt][r]{\@thefnmark\,}#1}
\makeatother 
\renewcommand{\figurename}{\small{Fig.}~}
\sectionfont{\sffamily\Large}
\subsectionfont{\normalsize}
\subsubsectionfont{\bf}
\setstretch{1.125} 
\setlength{\skip\footins}{0.8cm}
\setlength{\footnotesep}{0.25cm}
\setlength{\jot}{10pt}
\titlespacing*{\section}{0pt}{4pt}{4pt}
\titlespacing*{\subsection}{0pt}{15pt}{1pt}

\fancyfoot{}
\fancyfoot[LO,RE]{\vspace{-7.1pt}\includegraphics[height=9pt]{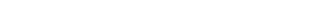}}
\fancyfoot[CO]{\vspace{-7.1pt}\hspace{13.2cm}\includegraphics{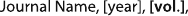}}
\fancyfoot[CE]{\vspace{-7.2pt}\hspace{-14.2cm}\includegraphics{head_foot/RF}}
\fancyfoot[RO]{\footnotesize{\sffamily{1--\pageref{LastPage} ~\textbar  \hspace{2pt}\thepage}}}
\fancyfoot[LE]{\footnotesize{\sffamily{\thepage~\textbar\hspace{3.45cm} 1--\pageref{LastPage}}}}
\fancyhead{}
\renewcommand{\headrulewidth}{0pt} 
\renewcommand{\footrulewidth}{0pt}
\setlength{\arrayrulewidth}{1pt}

\setlength{\columnsep}{6.5mm}
\setlength\bibsep{1pt}

\makeatletter 
\newlength{\figrulesep} 
\setlength{\figrulesep}{0.5\textfloatsep} 

\newcommand{\topfigrule}{\vspace*{-1pt}%
\noindent{\color{cream}\rule[-\figrulesep]{\columnwidth}{1.5pt}} }

\newcommand{\botfigrule}{\vspace*{-2pt}%
\noindent{\color{cream}\rule[\figrulesep]{\columnwidth}{1.5pt}} }

\newcommand{\dblfigrule}{\vspace*{-1pt}%
\noindent{\color{cream}\rule[-\figrulesep]{\textwidth}{1.5pt}} }

\makeatother

\twocolumn[
  \begin{@twocolumnfalse}
{\includegraphics[height=30pt]{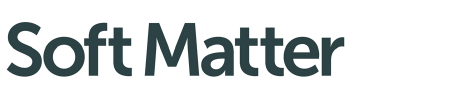}\hfill\raisebox{0pt}[0pt][0pt]{\includegraphics[height=55pt]{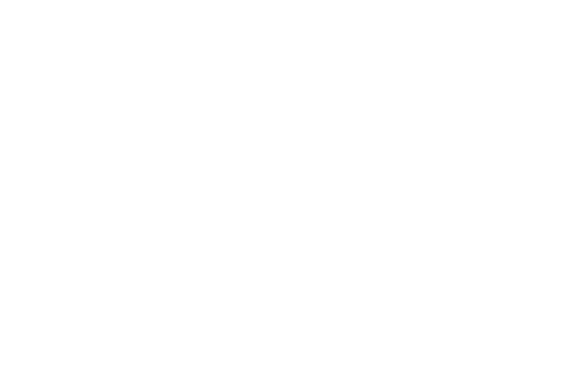}}\\[1ex]
\includegraphics[width=18.5cm]{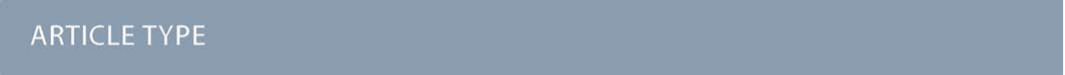}}\par
\vspace{1em}
\sffamily
\begin{tabular}{m{4.5cm} p{13.5cm} }

\includegraphics{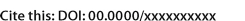} & \noindent\LARGE{Precisely controlled colloids: A playground for path-wise non-equilibrium physics}
\\
\vspace{0.3cm} & \vspace{0.3cm} \\

 & \noindent\large{Cai Dieball\textit{$^{a}$}, and Yasamin Mohebi-Satalsari\textit{$^{b}$}, and  Angel B. Zuccolotto-Bernez\textit{$^{b}$}, and Stefan U.~Egelhaaf\,\textit{$^{b}\dag$}, and Manuel A.~Escobedo-S\'anchez$^{\ast}$\textit{$^{b}$}, and 
 Alja\v{z} Godec$^{\ast}$\textit{$^{a}$}}\\

\includegraphics{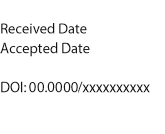} & \noindent\normalsize{
We investigate path-wise observables in experiments on driven colloids in a periodic light field to dissect selected intricate transport features, kinetics, and transition-path time statistics out of thermodynamic equilibrium. These observables directly reflect the properties of individual paths in contrast to the properties of an ensemble of particles, such as radial distribution functions or mean-squared displacements. In particular, we present two distinct albeit equivalent formulations of the underlying stochastic equation of motion, highlight their respective practical relevance, and show how to interchange between them. We discuss conceptually different notions of local velocities and interrogate one- and two-sided first-passage and transition-path time statistics in and out of equilibrium. \blue{Our results reiterate how} path-wise observables may be employed to systematically assess the quality of experimental data and demonstrate that, given sufficient control and sampling, one may quantitatively verify subtle theoretical predictions.
}\\

\end{tabular}

 \end{@twocolumnfalse} \vspace{0.6cm}

  ]

\renewcommand*\rmdefault{bch}\normalfont\upshape
\rmfamily
\section*{}
\vspace{-1cm}


\footnotetext{\textit{$^{a}$~Mathematical bioPhysics Group, Max Planck Institute for Multidisciplinary Sciences, 37077 G\"ottingen, Germany.; E-mail: agodec@mpinat.mpg.de}}
\footnotetext{\textit{$^{b}$~Condensed Matter Physics Laboratory, Heinrich Heine University D\"usseldorf, 40225, D\"usseldorf, Germany.; E-mail: escobedo@hhu.de}}




\section{Introduction}
Colloidal particles, due to their high susceptibility to external fields, can be precisely manipulated using light
\cite{Loewen2,Stefan4,Volpe,Ashkin,Grier,martinez2016brownian,Militaru,Zunke,Loos_PRX,Ibanez2024NP,Martinez2017SM,blickle2012realization}, electrical and magnetic fields  \cite{magnetic,Loewen2,magnetic3,magnetic4}, or microfluidic devices
\cite{fluidic}. This makes them an ideal platform to validate fundamental physical theories of soft matter with a high degree of accuracy. Over the years, substantial effort has been made in colloidal soft matter to explore various aspects of both stochastic dynamics \cite{Dhont1996,Naegele,Raphael,Raphael_2} and stochastic thermodynamics \cite{Sekimoto2010, Seifert2012RPP}. In fact, colloidal systems have always served as a paradigm for stochastic thermodynamics \cite{Sekimoto2010,Seifert2012RPP}, which generalizes the notion of thermodynamic observables to individual stochastic paths. Many fundamental kinetic and thermodynamic properties have been discovered and verified using colloids, including the statistics of work performed on \cite{Wang2002PRL,Imparato2007PRE} and heat dissipated by \cite{Imparato2007PRE} a driven colloid, detailed \cite{Wang2005PRE} and transient \cite{Carberry2004PRL} fluctuation theorems, realizations of heat engines \cite{blickle2012realization,martinez2016brownian,Martinez2017SM}, as well as first-passage \cite{Magazzu2019SM,Thorneywork2020SA,Zunke} and transition-path time statistics \cite{Zijlstra2020PRL,Neupane2012PRL,Gladrow2019NC,Satija2020PNASU,Sturzenegger2018NC}, to name a few. Notably, these are properties of individual stochastic paths, so-called functionals, which inherently provide much deeper insight into the underlying dynamics than their ensemble-average counterparts \cite{Majumdar2002PRL,Bel2005PRL,Eli_TA,Sabhapandit2006PRE,Lapolla2020PRR,Dieball2022PRL,Dieball2022PRR}. 

Despite decades of intensive research leading to numerous significant discoveries, the potential of driven and controlled colloids to reveal fundamental physical laws remains far from exhausted. On the one hand, this may be because advances in abstract theory (e.g., functional fluctuation relations \cite{Baiesi2009PRL,Dechant2020PNASU,Dieball2022PRL}, speed limits \cite{CSL_3,CSL_7,VanVu2023PRX,Arxiv_transport}, thermodynamic uncertainty relations \cite{Barato2015PRL,Horowitz2019NP,Macieszczak2018PRL,Koyuk2019PRL,Koyuk2020PRL,Dechant2021PRX,Massi,Fu2022PRE,Koyuk2022PRL,VanVu2023PRX,Dieball2023PRL}) do not so easily proliferate to experiments or require excellent statistics. On the other hand, reciprocal-space-based techniques (e.g., dynamic light scattering, neutron or X-ray scattering, Differential Dynamic Microscopy and variants) \cite{Cerbino2008,Kamal2024,Giavazzi_2014,Escobedo_S_nchez_2018} are only beginning to be considered in the theory of stochastic thermodynamics \cite{Arxiv_transport}. 

To go beyond the state-of-the-art in particle-tracking analysis we focus on path-wise observables that directly reflect properties of entire individual paths rather than properties of the probability density of an ensemble of particles, such as radial distribution functions or mean-squared displacements. \blue{Path-wise refers to those functionals of \emph{particular} realizations of trajectories that map entire trajectories $(X_\tau)_{0\le\tau\le t}$ or large parts of them to some $f[(X_\tau)_{0\le\tau\le t}]$, as opposed to observables depending only on the value at some fixed set of times, e.g., $f(X_{\tau_1},X_{\tau_2})$.}
A particular class of insightful path-wise observables is first-passage time\cite{Redner2001,Metzler2013}, which is the stochastic time it takes for a trajectory to reach a prescribed target (position) for the first time\blue{, e.g., $f[(X_\tau)_{0\le\tau\le t}]={\rm arg\,min}_{0\le\tau\le \infty}(X_\tau=a)$ for a target at $a$}. For example, first-passage time statistics have been shown to provide a deeper understanding of the origin of sub-diffusion than mean-squared displacements \cite{FPT}. Moreover, they can distinguish processes with equal transition probability densities \cite{Igor}, unravel the number of intermediate states\cite{Thorneywork2020SA}, and reveal fractal dynamics of colloids \cite{Zunke}. Related, albeit quite different, path-wise observables are transition-path times defined as the stochastic duration of successful transitions, whereby ``successful'' reflects that the particle does \emph{not} return to its original position before arriving at a predefined target point \cite{Berezhkovskii2006PRL,Dima_PNAS,Dima_Test,Hartich2021PRX}. Under typical conditions, transition-path times obey a surprising symmetry\cite{Berezhkovskii2006PRL}, and violations of this symmetry may be used to gain intriguing insights \cite{Gladrow2019NC}.

The ``inherent'' sensitivity of the selected observables is such that they can be systematically used to assess and critically verify the quality of experimental data. In this study, we conduct precise experiments on colloids driven through a periodic light field to investigate fundamental and complex aspects of transport,  along with first-passage and transition-path times in out-of-equilibrium conditions. \blue{Our results reiterate how} given sufficient control and sampling even the most subtle theoretical predictions may be verified quantitatively, which will hopefully \blue{reinforce the} motivation in the field of experimental soft matter to test and further increase the quality of experimental data.
 \subsection*{Structure of the paper}
In the first part, we present two different ways to write the underlying stochastic equation of motion on a trajectory-based level and show how to interchange between the two in theory and practice. 
We highlight the necessity of knowing both representations by connecting them to essential dynamic and thermodynamic properties. Based on these representations, we discuss conceptually different notions of local velocities, whose interrelations are a priori \emph{not} obvious. Next, we investigate first-passage time statistics for barrier-crossing events and show how these are linked to the local mean velocity (using Ref.~\cite{Reimann2001PRL}). Finally, we verify (and slightly extend to periodic systems) the transition-path time symmetry predicted in Refs.~\cite{Berezhkovskii2006PRL,Hartich2021PRX}. All of the aforementioned aspects are confirmed and supported by experimental data on both equilibrium (passive) and driven experiments. We conclude with a perspective on further research directions and open questions.

\section{Materials and Methods}
\label{Materials and Methods}
In this study, the applied potential is periodic in the $x$-direction and originates from the optical force resulting from the interaction of particles with a periodic light field with period \blue{$L=4.135\ \mu\text{m}$.}

\subsection{Sample preparation}\label{SamplePreparation}
\noindent We prepared a dilute colloidal dispersion containing polystyrene Sulfate latex particles of 1.5 $\mu$m radius with a polydispersity of 4 $\%$ (Thermo Fisher Scientific, batch number 1660463). The particles were suspended in ultra-pure water with a resistivity of 18.2 M$\Omega$cm (Purelabs Flex, Elga). The dispersion was confined to quasi-two-dimensional (2D) sample cells, which were assembled as follows: 2.3 $\mu$L of the dilute colloidal dispersion was placed on top of \blue{a 22X50 mm cover slip}, then \blue{a 22X22 mm cover slip} was carefully placed on top, and the slides were glued together using UV-curing glue (NOA61, Norland Products Inc.). We mounted the assembled cell on a microscope slide. To avoid \blue{the two glasses to get too close to each other (by capillary forces) and pinning the particles}, we have used the polydispersity of the sample, i.e., the larger sizes in the dispersion (particles of around 2 $\mu$m in radius), as spacers. All sample cells were left to reach equilibrium in a laboratory environment at $20\ ^{\circ}$C for two days. The area fraction was $\varphi_{a}\simeq1\ \%$.

\subsection{Experimental Setup}\label{Experimental Setup}
\begin{figure}[ht!]
\begin{centering}
\includegraphics[width=1.00\linewidth]{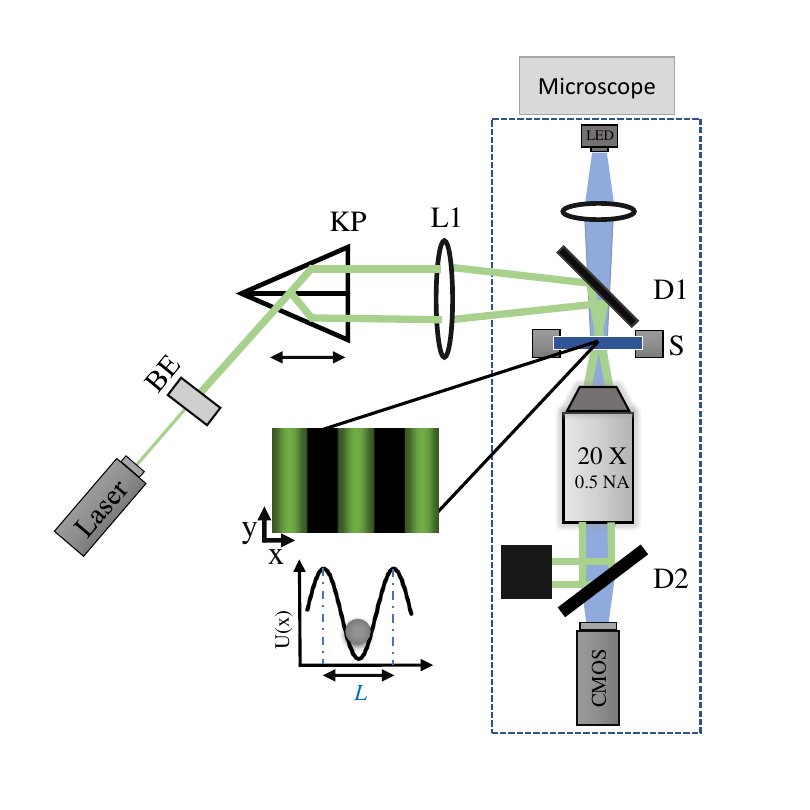}
\par\end{centering}
\protect\caption{\blue{Schematic representation of the experimental setup. A laser beam is expanded using a beam expander (BE) and directed to a Kösters prism (KP) to create two parallel beams. The laser beams are focused (using L1 and D1) in the sample plane to create the periodic light field. Dichroic mirrors D1 and D2 transmit the LED and reflect the laser wavelength. The sample cell is mounted on a piezo nanopositioner stage (S).}}
\label{fig:Setup}
\end{figure}

\blue{In this study, the periodic light field is created by the interference of two laser beams\cite{Capellman2017}. A laser beam (Cobolt 05-01 Samba 1.5 W) of 532 nm wavelength, is expanded and then split into two parallel beams using a Köster prism. \red{Using a lens and a dichroic mirror the beams are guided to interfere on the microscope sample plane}, creating a fringe pattern. The period ($L$, dark-bright fringe) can be changed by moving the Kösters prism position. The laser beams are removed from the image path (with a dichroic mirror) and a CMOS camera (Mako U-130B) is used as sensor to record images. {The sample cell sits on a piezo nanopositioner stage (Mad City Labs, Nano-BioS300), which is used to move the sample.}
A schematic representation of the experimental setup is shown in Fig.~\ref{fig:Setup}.
For details on the extraction of the period $L$ \red{and amplitude} of the potential, see Appedix \ref{sec:appendix}.}
\subsection{Optical Microscopy and Particle tracking}\label{Optical Microscopy and Particle tracking}
\blue{All experiments were performed with the colloidal particles under the effect of the periodic light field.}\\
\textbf {Equilibrium state (Not driven).}~We used an inverted bright-field microscope (Nikon, Ti-E) with a 20$\times$ objective (Nikon, Plan Flour, 0.5NA) and a light-emitting diode (Thorlabs M455L4) as an illumination source. The images were recorded with a CMOS camera (Mako U-130B) at a resolution of 1280$\times$1024 pixels utilizing the full field of view of the camera. The pixel pitch was 0.24 $\mu$m/px. The data acquisition was performed at 25 frames per second and an exposure time of 1 ms, with 90000 images per measurement. This corresponds to a total measurement duration of 1 hour. We conducted several experiments that reached a total collection of 1224 trajectories. We used modified MATLAB-based particle tracking routines based on those of D. Blair and E. Dufresne\cite{Matlabroutine} to accurately determine the position of the colloids. \blue{ Additionally, by following the Michalet algorithm\cite{Michalet2010}, we estimated the localization uncertainty to be $\pm2$ nm.} For a direct comparison to the driven state experiments, we treated the $1224$ trajectories as $2448$ trajectories with a duration of $30$ minutes each.\\

\noindent \textbf {Driven state.}~We used a piezo nanopositioner stage to \blue{apply a driving force to the colloids by dragging the sample cell through the potential.} The stage was programmed to move 139 $\mu$m  in 65500 steps, with a 40 ms pause between each step. Optimizing these values allowed us to drag the sample cell at a controlled velocity of $v_0$=$(0.053 \pm 0.002)\ \mu$m/s. We recorded images at 25 frames per second, each measurement containing 45000 images, for a total measurement time of 30 minutes.  In total, we acquired 1168 trajectories.

\section{Overdamped Langevin dynamics in a periodic drift field}\label{SamplingSchemes}
\subsection{Equilibrium dynamics}

The dynamics of colloidal particles are generically overdamped on the observed scales \cite{Dhont1996}. Thus, 
we consider the stochastic dynamics in one-dimensional space of a colloidal particle with position \blue{$X_t$} connected to a thermal bath with temperature $T$, governed by the overdamped Langevin equation
\blue{\begin{align}
\rmd X_t &= -\frac{D}{k_{\rm B}T}\partial_X U(X_t)\,\rmd t + \sqrt{2D}\,\rmd W_t
\label{equilibrium LE}\,,
\end{align}}
where \blue{$k_{\rm B}$ denotes the Boltzmann constant,} $D\propto T$ is the diffusion constant, $\rmd W_t$ is the increment of the fluctuating thermal force (Wiener process), and $U(x)$ is a potential, with units of energy. The force arising from the potential is $-\partial_x U(x)$ which results in the drift field $-D\partial_x U(x)/k_{\rm B}T$ (i.e., the fluctuation-dissipation theorem yields the mobility $\mu\equiv D/k_{\rm B}T$). 
On the level of probability densities of particle positions, the dynamics in Eq.~\eqref{equilibrium LE} is described by the Fokker-Planck equation \cite{Gardiner1985,Risken1989} $\partial_t p(x,t)=-\partial_x\{[-D\partial_xU(x)/k_BT-D\partial_x]p(x,t)\}$. However, we mainly use Eq.~\eqref{equilibrium LE} to stay closer to the path-wise description.\\ 

\begin{figure}[ht!]
    \begin{centering}
        \includegraphics[width=0.55\linewidth]{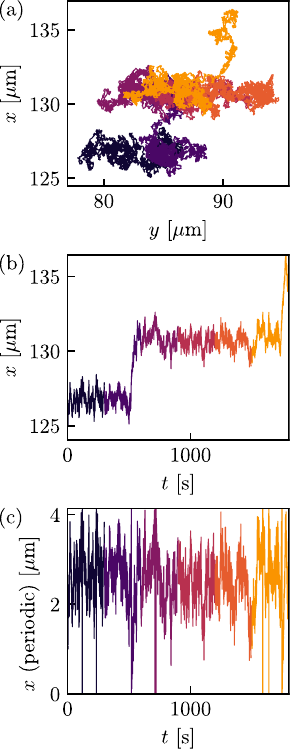}
    \par\end{centering}
    \protect\caption{Exemplary trajectory (a), and corresponding projection onto the $x$-axis (b,c) where time runs from dark to bright. (c) The period is always chosen as $L=4.135\ \mu$m.
    The dataset comprises 2448 such trajectories. 
    }
    \label{fig:EQ_trajs}
\end{figure}

As mentioned in Sec.~\ref{Materials and Methods}, in this study, the potential $U(x)$ is periodic in the $x$-direction with a period $L$. An experimental trajectory is shown in Fig.~\ref{fig:EQ_trajs}. \blue{Note that the motion in the $y$-direction will only correspond to a free Brownian motion, as there is no force coming from $U(x)$ in this direction. The latter is illustrated in Fig.~\ref{fig:EQ_trajs}a where the effect of $U(x)$ constrains the movement in the $x$-direction, contrary to the $y$-direction where it is allowed to move freely.} 

In Fig.~\ref{fig:EQ_trajs}b,c, we therefore focus on \blue{$X_t$}, where in (c) we show \blue{$x_t$ which is defined as $X_t$} projected onto a single period $[0,L)$\blue{, i.e., $x_t\equiv X_t\;{\rm mod}\;L$}. Accordingly, for $U(x)=U(x+L)$ we may view Eq.~\eqref{equilibrium LE} as $L$-periodic dynamics \blue{for $x_t$, i.e., $\rmd x_t = -{D}\partial_x U(x_t)\rmd t/{k_{\rm B}T} + \sqrt{2D}\,\rmd W_t$ with the definition $x_{t+\rmd t}\equiv (x_t+\rmd x_t)\;{\rm mod}\;L$ ensuring that $x_t$ remains confined to $[0,L)$}.
\blue{If projected on $[0,L)$, the} system settles into a Boltzmann equilibrium density $p_{\rm eq}(x)\propto\exp(-U(x)/k_{\rm B}T)$ \cite{Gardiner1985}.

More generally, not all systems settle into equilibrium steady states. For sufficiently confined systems and drifts without an explicit time dependence, a steady state is approached for $t\to\infty$. Similarly, for space-periodic dynamics treated as if they evolve in a single period, i.e., projected onto a single period, a steady state is approached for long times, assuming no explicit time dependence in the drift and diffusion. However, here, a stationary current emerges if the system is driven out of equilibrium by a non-conservative drift; see below.

\subsection{Driven dynamics}
\begin{figure}[h]
\begin{centering}
\includegraphics[width=0.55\linewidth]{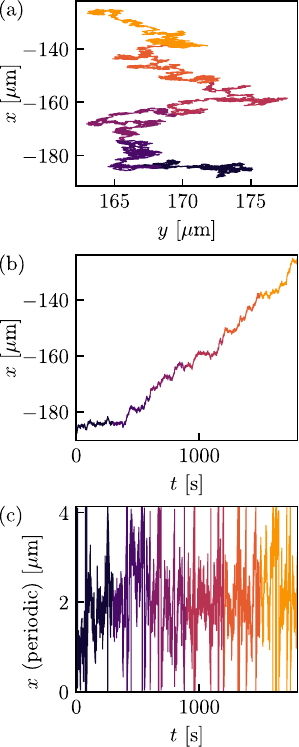}
\par\end{centering}
\protect\caption{\blue{Example of a trajectory} that is driven by $v_0$, see Eq.~\eqref{NESS LE 1}, (a), and corresponding projection onto the $x$-axis (b,c) where time runs from dark to bright. (c) The period is always chosen as $L=4.135\ \mu$m.
The dataset comprises 1168 such trajectories.
}
    \label{fig:NESS_trajs}
\end{figure}

While equilibrium dynamics are interesting and important, they are relatively well understood. However, they do not apply to irreversible (e.g., living) systems since these are inherently out of equilibrium, e.g., driven by non-conservative flows (i.e., shear) or ATP hydrolysis. Note that the theory presented here is \emph{not} new but is simply presented in a comprehensive manner and in a potentially new logical order to stay close to the experiment.
To address driven dynamics, we consider the simplest situation in which we add a constant \blue{bare velocity} $v_0$ in the $x$-direction to the Langevin equation \eqref{equilibrium LE}, yielding
\begin{align}
    \rmd x_t &= \left[-\frac{D}{k_BT}\partial_x U(x_t)+ v_0\right]\rmd t + \sqrt{2D}\,\,\rmd W_t\,. \label{NESS LE 1}
\end{align}
In the experiments, $v_0$ is introduced by {dragging} the sample cell with a constant velocity $-v_0$ along the periodic direction of $U(x)$ using the piezo nanopositioner stage. For a sample trajectory in this driven setting, see Fig.~\ref{fig:NESS_trajs}. Note that, as before, the motion in the $x$- and $y$-directions decouples and that the motion in the $y$-directions remains equilibrium Brownian motion only, see Fig.~\ref{fig:NESS_trajs}a. 

What can we say about the driven dynamics? First, the dynamics projected onto a single period, see Fig.~\ref{fig:NESS_trajs}c, still settles into a steady state---a non-equilibrium steady state (NESS)---with density $\ps(x)$, which, however, \emph{no} longer has a Boltzmann form. Instead, the shape of $\ps(x)$ is skewed in the direction of ${\rm sign}(v_0)$. 

The first question that arises is, given Eq.~\eqref{NESS LE 1}, what is $\ps(x)$ for a given bare velocity $v_0\ne 0$? The second question refers to the movement (i.e., probability current) of the driven particle; see Fig.~\ref{fig:NESS_trajs}b. This question is non-trivial, and compared to the bare velocity $v_0$, the particle is expected to slow down by the barriers of $U(x)$. Thus, one naturally wonders what theoretical velocity is implied by the equation of motion in Eq.~\eqref{NESS LE 1}.

Both questions can be answered by rewriting Eq.~\eqref{NESS LE 1} into an alternative form
\begin{align}
    \rmd x_t &= \left[D\partial_x \ln\ps(x_t)+\frac{\js}{\ps(x_t)}\right]\rmd t + \sqrt{2D}\,\,\rmd W_t\,. 
    \label{NESS LE 2}
\end{align}

In the following subsection, we will show how to derive it. First, we will discuss the properties and usefulness of this rewriting.  
Here, $\js$ is the steady-state probability current and $\nus(x)\equiv\js/\ps(x)$ is the ``local mean velocity'' \cite{Seifert2012RPP}\blue{. The latter }has to fulfill $\partial_x[\nus(x)\ps(x)]=0$ to ensure that $\ps(x)$ is indeed the steady state density\blue{, which is seen by requiring that $\partial_t p(x,t)=0$ for $p(x,t)=\ps(x)$, where $p(x,t)$ is the probability density goverend by the Fokker-Planck equation \cite{Gardiner1985,Risken1989},}
\begin{align}
\label{FPE}
        \partial_t p(x,t) &= -\partial_x\left\{[-D\partial_x U(x)/k_BT\blue{+v_0}-D\partial_x]p(x,t)\right\}\nonumber\\
        &= -\partial_x\left\{[D\partial_x \ln p(x,t)+\nu(x,t)-D\partial_x]p(x,t)\right\}\nonumber\\
        &= -\partial_x [\nu(x,t)p(x,t)]
        \,.
    \end{align}
    In one-dimensional space, this directly implies that there is a constant probability current $\js$ with $\nus(x)=\js/\ps(x)$. For equilibrium dynamics $v_0=0, \js=0$, and thus both presentations Eqs.~\eqref{NESS LE 1} and \eqref{NESS LE 2} agree, since \blue{for the equilibrium density $p_{\rm eq}(x)$ we have} $\partial_x \ln p_{\rm eq}(x)=-\partial_x U(x)/k_BT$. Thus, both can be seen as a direct extension of Eq.~\eqref{equilibrium LE}. 
    
Eq.~\eqref{NESS LE 2} is particularly insightful when considering microscopic reversibility (i.e., detailed balance) and its generalizations. Namely, if we let $G^{\nus}(x,t|x_0)$ denote the two-point conditional probability density of $x_t$ (which is the Green's function of the Fokker-Planck equation~\eqref{FPE}, i.e., $\partial_tG^{\nus}=-\partial_x[-D\partial_x\ln\ps(x)+\nus(x)]G^{\nus}$ with $G^{\nus}(x,0|x_0)=\delta(x-x_0)$), then detailed balance corresponds to
\begin{align}
G^{0}(x,t|x_0)\ps(x_0)=G^{0}(x_0,t|x)\ps(x).
\label{DB}
\end{align}
\blue{This} indeed holds if and only if $\nus(x)=0$ (i.e., $\js=0$)\blue{. In contrast,} the generalization of Eq.~\eqref{DB}--the so-called ``dual reversal'' symmetry--corresponds to \cite{Dieball2022PRR}
\begin{align}
G^{\nus}(x,t|x_0)\ps(x_0)=G^{-\nus}(x_0,t|x)\ps(x).
\label{Dual}
\end{align}
Note that the dual reversal symmetry Eq.~\eqref{Dual} and symmetries of dynamical functionals of $x_t$ (see Ref.~\cite{Dieball2022PRR}) are only applicable to the steady-state local mean velocity $\nus(x)$ in Eq.~\eqref{NESS LE 2} but not to the bare velocity $v_0$ in Eq.~\eqref{NESS LE 1}.  On the other hand, the potential $U(x)$ that enters thermodynamic potentials only appears in the first form \eqref{NESS LE 1}.
The main difference between the representations Eq.~\eqref{NESS LE 1} and Eq.~\eqref{NESS LE 2} is that the additional term $\nus(x)=\js/\ps(x)$ (unlike $v_0$) does \emph{not} alter the steady-state density $\ps(x)$, irrespective of the value of the constant steady-state current $\js$. 

When discussing the velocity of the driven system, we must note that
$\dot x_t$ does \emph{not} exist \blue{in a mathematical sense} for overdamped motion \blue{as in Eqs.~\eqref{NESS LE 2} and \eqref{NESS LE 1}}  [technically ${\rm Prob}(\abs{\dot x_t}<C)=0$ for all $C<\infty$] and that there are, in fact, different notions of velocity. The bare velocity $v_0$ and the drift velocity \blue{$v_{\rm drift}(x)$} (where $\E{\cdot|x_\tau=x}$ denotes the average conditioned on $x_\tau=x$ \blue{for any $\tau\in[0,t]$}),
\begin{align}
    \frac{\E{\rmd x_\tau|x_\tau=x}}{\rmd t} = \blue{v_{\rm drift}(x)} \equiv D\partial_x \ln\ps(x)+\nus(x) = -\partial_x U(x)/k_BT + v_0\,,
    \label{drift velocity}
\end{align}
are directly accessible from Eq.~\eqref{NESS LE 1}\blue{. In contrast to $v_0$ and $v_{\rm drift}(x)$, }the local mean velocity $\nus(x)$ and the mean velocity $\bar v$,
\begin{align}
    \bar v &\equiv \frac{\Es{\rmd x_\tau}}{\rmd\tau}=\int_0^L\rmd x [D\partial_x \ln\ps(x)+\nus(x)]\ps(x)\nonumber\\
    &=\int_0^L\rmd x [D\partial_x\ps(x)+\js]
    =L\js\,, \label{mean velocity definition}
\end{align}
[where $\Es{\cdot}$ denotes the expectation over paths \blue{generated by Eq.}~\eqref{NESS LE 1} evolving from $\ps(x_0)$],
are obtained from the second form of the equation of motion~\eqref{NESS LE 2}. \blue{Note that for initial conditions other than $\ps(x_0)$, a mean velocity defined as $\E{\rmd x_\tau}/\rmd\tau$ would not be constant in time but only relaxes towards the above $\bar v$ as $t\to\infty$. Moreover, note} that for a periodic $U(x)$ (with known period $L$), the mean velocity is equivalently characterized by the following expressions \blue{(recall that $X_\tau$ is the full dynamics and $x_\tau\equiv X_\tau\;{\rm mod}\;L$)},
\begin{align}
    \bar v &\equiv \frac{\Es{\rmd x_\tau}}{\rmd\tau} = \Es{\blue{v_{\rm drift}(x_\tau)}} = \Es{\nus(x_\tau)} = L\js\nonumber\\
    &=\frac{\Es{\blue{X_{t_2}-X_{t_1}}}}{t_2-t_1} = \frac L{\Es{\tau(x\to x+L)}}\,,
    \label{mean velocity}
\end{align}
for any times $t_1$, $t_2$ and for any $x$, and  we used  $\Es{\partial_x\ln\ps(x_\tau)} =\int_0^L\frac{\partial_x\ps(x)}{\ps(x)}\ps(x)\rmd x = \ps(L)-\ps(0) =0$ to show $\Es{\blue{v_{\rm drift}(x_\tau)}}= \Es{\nus(x_\tau)}$ as well as $\Es{\blue{X_{t_2}-X_{t_1}}} = \Es{\int_{\tau=t_1}^{\tau=t_2}1\circ\rmd \blue{X_\tau}}=\int_0^t\rmd\tau\int_0^L\rmd x 1\js=tL\js$ (see, e.g., Ref.~\cite{Dieball2022PRR} for details). The term $\tau(x\to x+L)$ in Eq.~\eqref{mean velocity} denotes the first-passage time from $x$ to $x+L$, i.e., the (stochastic) time that a \blue{trajectory $(X_\tau)_{0\le\tau\le t}$} starting at position $x$ takes to reach position $x+L$ for the first time, see also Fig.~\ref{fig:FPT_transition_time}. The last equality in Eq.~\eqref{mean velocity} is shown in Ref.~\cite{Reimann2001PRL} and will be revisited later in this work. Before we demonstrate and verify the different notions of velocities from the experimental data, we first need to find a way to swap between the representations, that is,  from Eq.~\eqref{NESS LE 1} to Eq.~\eqref{NESS LE 2} and vice versa.

\subsection{Swapping representations:~From Eq.~\eqref{NESS LE 1} to Eq.~\eqref{NESS LE 2}}
On a general note, we emphasize that determining Eq.~\eqref{NESS LE 2} analytically in a high-dimensional space is generally \emph{not} feasible (i.e., it requires solving the stationary Fokker-Planck equation $\partial_t p(x,t)=0$; in practice, one would need to simulate long trajectories and estimate $\ps(x)$ as histograms).
However, in the given one-dimensional scenario with constant bare velocity $v_0$ there is a way to obtain $\ps(x)$ and, consequently, Eq.~\eqref{NESS LE 2} from Eq.~\eqref{NESS LE 1}, see Ref.~\cite{Gardiner1985}, which we follow here.

\begin{figure}[h]
\begin{centering}
\includegraphics[width=0.9\linewidth]{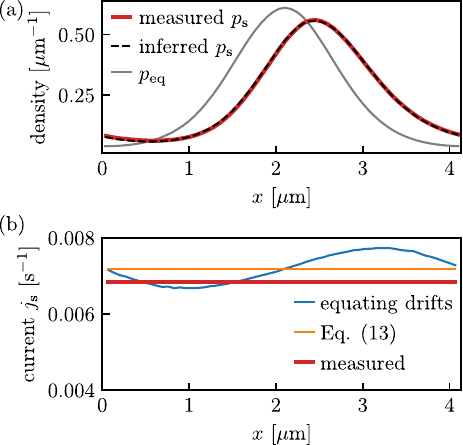}
\par\end{centering}
\protect\caption{(a) $\ps(x)$ in Eq.~\eqref{1d result ps(x)} inferred from
      equilibrium (passive) experiments \red{and using the value $v_0=0.053$ $\mu$m/s of the experimentally fixed velocity}, where we \blue{deduced} $U(x)/k_{B}T$
      up to a constant offset from $\ln p_{\rm eq}$, \blue{where $p_{\rm eq}$}
      was estimated as a histogram \blue{of the equilibrium dynamics }with $60$ bins in $[0,L)$ (setting a limited
      spatial resolution for the rest of the
      analysis).
      Note that for our experimental setting,
        $U$ is technically known but involves a convolution with a
        Bessel function \cite{65,Bewerunge_A} and is prone to
        experimental inaccuracies. We thus prefer to infer $U$
        directly from the equilibrium data.
        The diffusion coefficient $D=0.04$ $\mu$m$^2$/s was fitted from $\var(\rmd x_\tau)/2\rmd\tau$ averaged over $[0,t]$ and over all trajectories, see Eq.~\eqref{infer D}. \blue{The equilibrium density $p_{\rm eq}$ is only shown for comparison and we stress that it, as expected, does not agree with $\ps(x)$.} (b) Probability current inferred from the equilibrium data using the value $v_0=0.053$ $\mu$m/s of the experimentally prescribed velocity as described in Eq.~\eqref{1d equation for js} and below. The comparison to the measured value in the driven data, see Eq.~\eqref{measured js}, shows slight deviations.}
    \label{fig:inferred_ps_js}
\end{figure}

Define the auxiliary function $\psi(x)$ (here $0$ is the left side of the periodic interval, and $x\in[0,L]$) as
\begin{align}
\psi(x)\equiv\exp\left[\frac1D\int_0^x \blue{v_{\rm drift}(x')}\rmd x'\right]=\exp\left[\frac{U(0)-U(x)}{k_BT}+\frac{v_0x}D\right]\,.
\label{1d definition psi} 
\end{align}
The result for $\ps(x)$ 
is
\begin{align}
\ps(x)&=\ps(0)\psi(x)\left(1-r(x)\left[1-\psi(L)^{-1}\right]\right)\,,\label{1d result ps(x)} 
\end{align}
where we introduced
\begin{align}
r(x)&\equiv\frac{\int_0^x\psi(x')^{-1}\rmd x'}{\int_0^L\psi(x')^{-1}\rmd x'}\in[0,1]\quad{\rm for}\ x\in[0,L]\,,\nonumber\\
\ps(0)&=\left[\int_0^L\psi(x)\left(1-r(x)\left[1-\frac1{\psi(L)}\right]\right)\rmd x\right]^{-1}\,.\label{aux_1d} 
\end{align}
As a consistency check, we note that for $v_0=0$ we should recover the Boltzmann distribution, and indeed we find $\psi(x)=\exp({[U(0)-U(x)]}/{k_BT})$ such that $\psi(L)=e^0=1$ and $\ps(x)/\ps(0)=\psi(x)$ provides it. Additionally, in Fig.~\ref{fig:inferred_ps_js}a, we also verify Eq.~\eqref{1d result ps(x)} \blue{as its resulting  $p_s$ (dashed line) overlaps with the $p_s$ measured from a histogram of the driven dynamics (red line).}
with experimental data.
To obtain the representation in Eq.~\eqref{NESS LE 2} in full, we are also required to fit the constant $\js=\nus(x)\ps(x)$. This may be done either from
\begin{align}
j_{\rm s}=D\ps(0)\frac{\psi(0)^{-1}-\psi(L)^{-1}}{\int_0^L\psi(x')^{-1}\rmd x'}\,,\label{1d equation for js} 
\end{align}
or by equating $-D\partial_x U(x)/k_BT+v_0$ in Eq.~\eqref{NESS LE 1} with $D\partial_x\ln\ps(x)+\frac{j_{\rm s}}{\ps(x)}$ in Eq.~\eqref{NESS LE 2}. 
With $\ps$ and $\js$ (Eqs.~\eqref{1d result ps(x)} and
\eqref{1d equation for js}) we can finally transform Eq.~\eqref{NESS
  LE 1} into Eq.~\eqref{NESS LE 2}, as done for the experimental data
in Fig.~\ref{fig:inferred_ps_js}. \blue{We stress again that this transformation was only analytically feasible since we deal with a one-dimensional system and a constant bare velocity $v_0$.} While the inferred density
  $\ps$($x$) in Fig.~\ref{fig:inferred_ps_js}a fits the measurement very
  well, there are slight deviations between the current $\js$ \blue{in Fig.~\ref{fig:inferred_ps_js}b
  inferred from the equilibrium measurements using the input value of
  $v_0$ (orange line) }compared to the driven measurement \blue{(red line) or to inferring the current by equating the drifts in Eqs.~\eqref{NESS LE 1} and \eqref{NESS LE 2} using the input $v_0$ and $p_{\rm eq}(x)$ and $\ps(x)$ as in Fig.~\ref{fig:inferred_ps_js}a (blue line in Fig.~\ref{fig:inferred_ps_js}b; }note that the blue line
  should, in principle, be constant). We speculate that the deviations
  may be due to very slight imperfections in the periodicity $L$ that become amplified since 
\blue{the field of view comprises many periods (about $60L$, see Fig.~\ref{fig:appendix} in the appendix).}
\subsection{Swapping representations:~From Eq.~\eqref{NESS LE 2} to Eq.~\eqref{NESS LE 1}}
For completeness, we also consider the reversed mapping, where we measure the driven data and want to infer the underlying potential $U(x)$ and bare velocity $v_0$. Note that this is only possible if we know that the driving arises purely from a \emph{constant} drift velocity $v_0$, and we can only infer $U(x)$ up to a constant. Knowing $U(x)$ is very relevant for stochastic thermodynamics (i.e., for determining free energy, internal energy, and work, but not the heat and entropy production or the dynamics). In the present case, transforming in the reverse direction mainly serves as a consistency check.

\begin{figure}[h]
\begin{centering}
\includegraphics[width=0.9\linewidth]{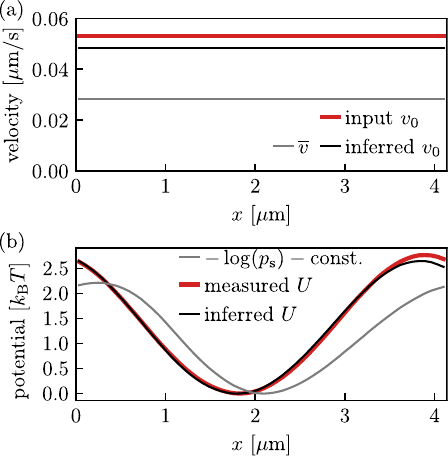}
\par\end{centering}
\protect\caption{(a) Inferred $v_0$ from driven experimental data compared to the desired input value $v_0$. \blue{Though there are some statistical deviations in the values for $v_0$ (related to Fig.~\ref{fig:inferred_ps_js}b), both obtained values for $v_0$ are well distinguishable from the mean velocity $\overline v$ (grey line)}. 
    (b) The potential $U(x)$ inferred from the driven data \red{via Eq.~\eqref{infer U} (black line)} matches the potential measured from the equilibrium data \red{projected on one period $[0,L)$ using $U(x)=-\ln(p_{\rm eq}(x))-$const. (red line)} quite well. \red{The barrier height (difference between minimum and maximum) is $2.76$ $k_{\rm B}T$ in the red line and $2.65$ $k_{\rm B}T$ in the black line, respectively.} It deviates from $-\log(\ps)$ (grey line) since the driven data set does not obey a Boltzmann density [see Eqs.~\eqref{NESS LE 1} and \eqref{NESS LE 2}].}
    \label{fig:inferred_U_v0}
\end{figure}

As for $p_{\rm eq}(x)$ in the equilibrium setting before, we estimate $\ps(x)$ from an ensemble of NESS trajectories as a histogram with $60$ bins, and the diffusion coefficient $D$ from the short-time fluctuations 
\begin{align}
    D &=\frac1t\int_0^t\rmd\tau\frac{\Es{\rmd x_\tau^2}}{2\rmd\tau}
    =\frac1{2t}\int_0^t\Es{\rmd x_\tau^2}\,.
    \label{infer D}
\end{align}
It turns out (at least from our trajectories length) that the easiest and most reliable way to obtain $\js$ from NESS trajectories appears to be to use [see Eq.~\eqref{mean velocity}]
\begin{align}
    \js = \frac{\Es{\blue{X_t-X_0}}}{Lt}\,,
    \label{measured js}
\end{align}
which \blue{together with $\ps(x)$} yields the form~\eqref{NESS LE 2}. To transform into Eq.~\eqref{NESS LE 1} we compute the bare velocity $v_0$ [comparing Eqs.~\eqref{NESS LE 1} and \eqref{NESS LE 2}; note that $U(x)$ is \emph{not} yet known at this point, but it drops out upon integration] as
\begin{align}
v_0
&=\frac1L\int_0^L\rmd x\left[D\partial_x \ln\ps(x)+\frac{\js}{\ps(x)}-D\partial_xU(x)/k_BT\right]
=\frac1L\int_0^L\rmd x\nus(x)\nonumber\\
&=\frac{\js}L\int_0^L\rmd x\frac{1}{\ps(x)}\,.\label{infer v0}
\end{align}
Now, we can obtain the potential (up to an additive constant) from its derivative [obtained by equating drift terms in Eqs.~\eqref{NESS LE 1} and \eqref{NESS LE 2}]
\begin{align}
    \partial_xU(x) &= -\frac{k_BT}D\left[\frac{D\partial_x\ps(x)+\js}{\ps(x)}-v_0\right]\,.\label{infer U}
\end{align}
This way, one can transform back from Eq.~\eqref{NESS LE 2} to
Eq.~\eqref{NESS LE 1}. In practice, this allows us to check whether the
driving $v_0$ introduced in the experiment is what we expected
(i.e., it serves as another consistency check in addition to
Fig.~\ref{fig:inferred_ps_js}). The potential $U(x)$ [we set the
additive constant to zero, i.e., we chose $\min U(x)=0$] and $v_0$
computed this way are compared to the measured equilibrium \blue{data (i.e., $v_0=0$) and the results are shown in Fig.~\ref{fig:inferred_U_v0}}. As
before, we observed slight deviations in velocity and current, probably
connected to slightly washed-out barriers
\blue{over the large field of view in the experiment (a tiny
error in the value of $L$ can become relevant when projecting on $[0,L)$ via $x_\tau= X_t\;{\rm mod}\;L$ since the range of $X_\tau$ comprises a range of about $60L$, see Fig.~\ref{fig:appendix} in the Appendix.} 
\blue{However, in general} the consistency check gives excellent results.
\noindent 
\section{Different notions of velocity}
\begin{figure}[h]
\begin{centering}
\includegraphics[width=0.90\linewidth]{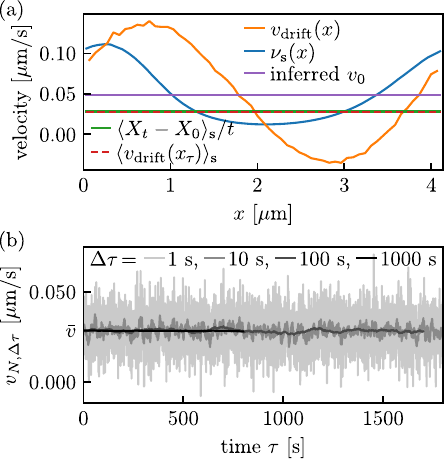}
\par\end{centering}
\protect\caption{
(a) Different space-dependent and constant velocities in the NESS. See Eq.~\eqref{mean velocity} for different representations of the mean velocity $\overline v$. In equilibrium, only $\blue{v_{\rm drift}(x)}$ deviates from zero. Quantities denoted by averages over $x_\tau$ are independent of $\tau$ in the steady state and are \blue{here, for practical reasons to improve statistics,} averaged over all $\tau\in[0,t]$. (b) While $\overline v=\Es{x_{\tau+\Delta\tau}-x_\tau}/\Delta\tau$ for all $\tau$ and $\Delta\tau$, see Eq.~\eqref{mean velocity}, the approximation $\overline v_{N,\Delta\tau}(\tau)=\frac1N\sum_{i=1}^N({x^i_{\tau+\Delta\tau}-x^i_\tau})/{\Delta\tau}$ of this quantity over the finite number of trajectories $N=1168$ as a function of $\tau$, as expected, exhibits large fluctuations for small $\Delta\tau$.}
    \label{fig:velocities}
\end{figure}

To provide more insight into the different notions of velocity in our system, we compute the bare velocity $v_0$, the drift velocity $\blue{v_{\rm drift}(x)}$ in Eq.~\eqref{drift velocity}, and the mean velocity $\bar{v}\blue{={\Es{X_t-X_0}}/{t}}$ in Eq.~\eqref{mean velocity} from the measured trajectories of the driven system. The results are shown in Fig.~\ref{fig:velocities}. 

As mentioned above, two constants \blue{constitute a notion of} velocity: the bare velocity $v_0$ and the mean velocity $\bar v$, see Eq.~\eqref{mean velocity}. In addition, there are also local, $x$-dependent notions like the local mean velocity $\nus(x)$, as well as the drift velocity $\blue{v_{\rm drift}(x)}=\E{\rmd x_t|x_t=x}/\rmd t$, see Eq.~\eqref{drift velocity}. In equilibrium, only the drift velocity $\blue{v_{\rm drift}(x)}$ deviates from $0$. The mean velocity $\bar v$ can be computed in different ways; see Eq.~\eqref{mean velocity}, some of which are also equivalent from a practical/numerical point of view (depending on how we choose to infer $\js$). Note that generally, $\nu_s(x)\ne\Es{\nu_s(x_t)}\ne \blue{v_{\rm drift}(x)}\ne\Es{\blue{v_{\rm drift}(x_t)}}$, that is, the different kinds of velocities indeed have a quite strikingly different meaning, which must be taken into account when characterizing and comparing transport properties.  

\section{First-passage times}
\begin{figure}
\begin{centering}
\includegraphics[width=0.90\linewidth]{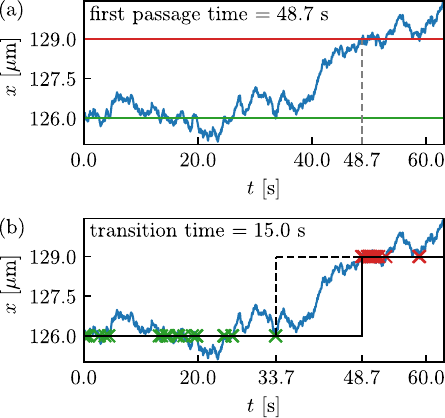}
\par\end{centering}
\protect\caption{Pictorial definition of the first-passage time (a) and transition time (b) to go from $x=126\ \mu$m to $x=129\ \mu$m. (a) For the trajectory starting at $x=126\ \mu$m at $t=0$ s, the first time it reaches $x=126\ \mu$m is after $46.7$ s \blue{which} is the value of the first-passage time for this trajectory snippet. (b) A transition time from $x=A$ to $x=B$ (will be considered in Sec.~\ref{sec:transition times}; here $A=126\ \mu$m and $B=129\ \mu$m) is the time from hitting $A$ for the last time before reaching $B$. In this example one transition $A\to B$ happens, which takes $15.0$ s.}
    \label{fig:FPT_transition_time}
\end{figure}
In this section, we evaluate the mean first-passage times, that is, the average time it takes for a particle to reach a certain point [see Fig.~\ref{fig:FPT_transition_time}; in our case, \blue{the average time} to cross the barrier in the potential $U(x)$] for the first time given some initial condition \cite{Redner2001}, using experimental data. The first-passage time for barrier crossing is an insightful observable, important, \blue{for instance, in the study of chemical reactions kinetics} \cite{Haenggi1990RMP}. Perhaps more important in the present context, first-passage observables are generally much more sensitive dynamics indicators than propagators, currents, or mean-squared particle displacement \cite{FPT,Igor,Zunke}. \blue{As such, they are }particularly suitable for \blue{critically assessing whether inferred properties of the dynamics (e.g., the parameters in the equations of motion) are appropriate}. In particular, first-passage observables may distinguish between processes with identical transition probability densities \cite{Igor}, and, in contrast to mean-squared particle displacements, provide insight into the microscopic origin of anomalous dispersion in complex media \cite{FPT,Zunke}.   

Here we focus on two particular first-passage observables, namely the mean-first passage to transverse a distance equal to the period length $L$, $\langle \tau(x\to x+L)\rangle$ and the two-sided mean exit time from the interval $[x-L,x+L]$, $\langle \tau_{\rm E}(x\pm L)\rangle$. Note that we do \emph{not} consider periodic boundary conditions here, \blue{i.e., we consider trajectories $(X_\tau)_{0\le\tau\le t}$ which evolve} on the entire \blue{space and not only on $[0,L)$}. \blue{The chosen first-passage observables have} two advantages. First, for a periodic potential $U(x)$, they are independent of the initial position $x$ \cite{Reimann2001PRL}, making them very practical to infer from experimental data. Second, the mean of its one-sided version, $\Es{\tau(x\to x+L)}$ is directly related to the mean velocity $\Es{\tau(x\to x+L)}=L/\bar{v}$, see Eq.~\eqref{mean velocity}.

To recapitulate where the independence comes from, consider $\tau(0\to L)$ and $\tau(x\to x+L)$ for some $x\in[0,L]$. Since we have one-dimensional trajectories, $x$ is necessarily crossed in the transition $0\to L$, and by the renewal theorem \cite{Siegert,Keilson} (note that we are dealing with a time-homogeneous Markov process with continuous paths) we immediately have statistical equality on the level of $\tau$ for $\tau(0\to L)=\tau(0\to x)+\tau(x\to L)$. Moreover, by periodicity we obtain $\tau(0\to x)+\tau(x\to L)=\tau(L\to x+L)+\tau(x\to L)=\tau(x\to L)+\tau(L\to x+L)=\tau(x\to x+L)$, therefore $\tau(x\to x+L)$ (and therefore $\langle \tau(x\to x+L)\rangle$) is independent of $x$, and similarly for $ \tau_{\rm E}(x\pm L)$. We will exploit this in the notation and drop the dependence on $x$, i.e., $\langle \tau(x\to x+L)\rangle\to \langle \tau(L)\rangle$.  

The $x$-invariance helps to check the data for short-time bias due to under-sampling (see Appendix of Ref.~\cite{Zunke}). Namely, whenever we infer $\tau$ from finite trajectories, say of duration $t$, starting at an arbitrary point $x_0$, we estimate a conditional mean  first-passage/exit time, $\langle \tau(L)|\tau(L)<t\rangle$~\cite{Zunke}, i.e.,
\begin{align}
\langle \tau(L)|\tau(L)<t\rangle\equiv \frac{\int_0^t\tau\wp_L(\tau)\rmd \tau}{\int_0^t\wp_L(\tau)\rmd \tau},
\label{cMFPT}    
\end{align}
where $\wp_L(\tau)$ is the probability density of $\tau(L)$, and the same holds for the two-sided exit time $\langle \tau_{\rm E}(L)|\tau_{\rm E}(L)<t\rangle$. \blue{Obviously, $\lim_{t\to\infty}\langle \tau(L)|\tau(L)<t\rangle=\langle \tau(L)\rangle$. For a periodic $U(x)$ in the equilibrium setting $v_0=0$, $\langle \tau(L)\rangle$ unlike the exit time $\langle \tau_{\rm E}(L)\rangle<\infty$  is infinite, $\langle \tau(L)\rangle=\infty$ since the particle can escape to $-\infty$ and thereby may \emph{not} reach the target at any finite time.  However, by the fundamental property of Brownian motion, in the equilibrium setting \emph{all} trajectories eventually hit the target.} Moreover, in the driven setting, we consider a bias towards the target, such that almost all trajectories reach the target in a finite time. Hence, we expect $\Es{\tau(L)}<\infty$ \cite{Redner2001} (this would, of course, not be the case of a particle biased away from the target). 

\begin{figure}
\begin{centering}
\includegraphics[width=0.90\linewidth]{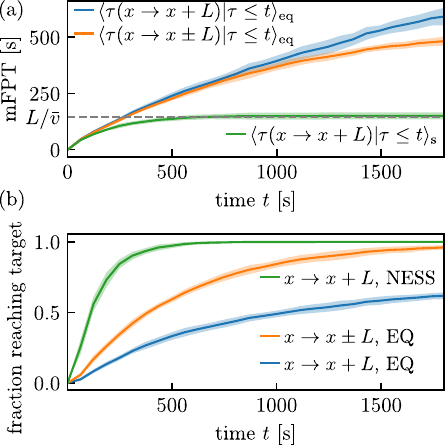}
\par\end{centering}
\protect\caption{(a) Conditional mean first-passage times (mFPT) and (b) fraction of trajectories where the barrier crossing event was realized conditioned on times until $t$ (i.e., we cut the data at time $t$ to mimic shorter measurements). The green line approaches $148.8$ s, which roughly agrees with the grey line at $146.4$ s, confirming the last equality in Eq.~\eqref{mean velocity}. Shaded regions show the sample standard deviation over the 11 (equilibrium ``EQ'') or 13 (NESS) data sets.}
    \label{fig:mFPT}
\end{figure}

Any substantial deviations between the conditional and unconditioned first-passage (and exit times, respectively) reflect that a significant fraction of trajectories did not yet cross the barrier to the right of the initial condition and that the estimated mean first-passage time is statistically unreliable. The manifestation of this short-time bias in the equilibrium and driven setting is demonstrated in Fig.~\ref{fig:mFPT}, where we mimic the effect of a progressively \blue{larger} measurement time $t$ in the experiment by disregarding long-time data. We find substantial effects of a finite duration of trajectories. In the equilibrium case, both the mean first-passage and the exit time increase with the duration $t$ of trajectories, while in the driven case, the mean first-passage time is progressively approaching the predicted value. 
Based on the Langevin dynamics in Eqs.~\eqref{NESS LE 1} or
\eqref{NESS LE 2}, the one-sided mean
first-passage time (blue line) is known to diverge as $t\to\infty$  in the equilibrium setting
[see, e.g., $\overline v\to0$ in Eq.~\eqref{mean velocity}], while the
two-sided version (orange line) must converge.  
In the non-equilibrium setting (green line), the mean first-passage time must converge, and we observe a quantitative agreement with the theoretical prediction $L/\bar{v}$ in Eq.~\eqref{mean velocity} (grey line), where the equality $\Es{\tau(L)}=L/\bar{v}$ is approached for large $t$. If the deviation of the green and grey dashed line in Fig.~\ref{fig:mFPT}a, does not (approximately) vanish. We could immediately infer that the measurement time is too short. Equivalently, the fraction of successful trajectories (see Fig.~\ref{fig:mFPT}b) ought to converge to $1$ for the exit from the interval and generally for the driven setting (and asymptotically approach $1$ for the one-sided first-passage in the equilibrium setting) if we are to infer a reliable estimate for $\langle\tau(L)\rangle$. However, note that the additional test with the mean velocity $\bar{v}$ may still be meaningful even if the fraction is already one (e.g., if one has a limited number of trajectories available or if one puts in the requirement of crossing the barrier as an additional condition).

Note that a short-time bias is, in fact, typical. \blue{Specifically,} whenever not (almost) \emph{all} observed trajectories reach the target, the estimated mean first-passage time (or its inverse, the rate) will suffer from this bias. \blue{This occurs because} the first-passage and the exit time are controlled (essentially dominated) by the long-time behaviour of $\wp(\tau)$ \cite{Redner2001,Olivier,A_PRX}. 

\section{Symmetry of transition-path times}\label{sec:transition times}
An essential and closely related, but fundamentally quite different, concept to first-passage times is \emph{transition-path times} \cite{Berezhkovskii2006PRL,Dima_Test,Dima_PNAS,Hartich2021PRX,JPCL_Dima}. The transition-path time $\mathcal{T}(A\to B)$ from a point $A$ to a point $B$ in (here considered in one-dimensional space) is defined as the time span between the last time that a trajectory $(x_\tau)_{\tau\ge0}$ hits $A$ before hitting $B$ for the first time, see Fig.~\ref{fig:FPT_transition_time}b. This observable is deeply related to the concept of mile-stoning (see, e.g., Refs.~ \cite{Elber,Sch_tte_2011,Hartich2021PRX,David_PRR,Blom2024PNASU}) and Fig.~\ref{fig:FPT_transition_time}b, where the crosses indicate hits of the milestones). The important difference to first-passage times $\tau(A\to B)$ is thus that the transition time $\mathcal{T}(A\to B)$ does not contain the dwell time around $A$ (see, e.g., Ref.~\cite{Hartich2021PRX} for details). A practical difference for our analysis is that we had $0$ or $1$ samples per trajectory for the first-passage time (since we always started at $t=0$) while a single trajectory may contain several transitions $A\to B$ and $B\to A$, see Fig.~\ref{fig:FPT_transition_time}. Moreover, we again consider the periodicity in space for the transition times.

Transition-path times encode subtle and important information about violations of the Markov property \cite{Dima_Test,Hartich2021PRX,Gladrow2019NC} and are essential for consistent thermodynamics of non-Markovian processes \cite{Hartich2021PRX,David_PRR,Blom2024PNASU,Gladrow2019NC}. A particularly subtle, and at first glance surprising, property of transition-path times is the forward-backward symmetry for reversible Markov dynamics discovered in Ref.~\cite{Berezhkovskii2006PRL} (for extensions, see Refs.~\cite{Berezhkovskii2019JCP,Satija2020PNASU,Hartich2021PRX,Ryabov2019JPCC}). It states that for a Markov process obeying detailed balance, we have the equality in time-distribution of transition paths, $p(\mathcal{T}(A\to B))=p(\mathcal{T}(B\to A))$. The surprising aspect is that if $A$ is located in a potential minimum and $B$ on a potential barrier, the duration of transition paths ``uphill'' and ``downhill'' are statistically identical for any potential.
Note that this symmetry does \emph{not} hold for first-passage times (since the dwell times in $A$ and $B$ can be arbitrarily different). 

\begin{figure}[ht!]
\begin{centering}
\includegraphics[width=0.95\linewidth]{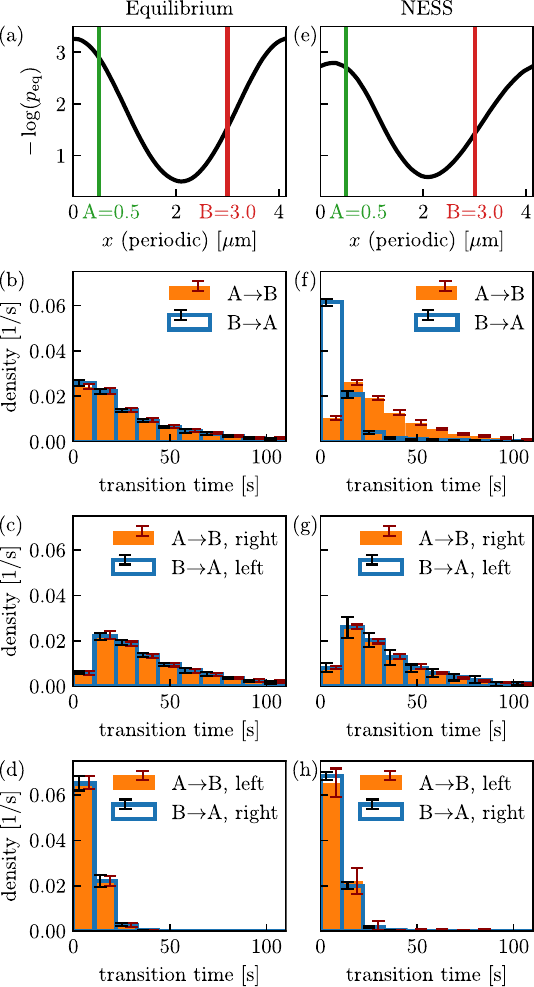}
\par\end{centering}
\protect\caption{
(a) Periodic potential for the equilibrium data and chosen milestones $A$ and $B$. (b) Probability of transition times for A$\to$B and B$\to$A, respectively. Error bars show the sample standard deviation over the different data sets. 
    (c, d) As in (b) but resolving for the direction of transitions. For both A$\to$B and \blue{B$\to$A}, about $70\ \%$ of the transitions belong to panel (c) and $30\ \%$ to panel (d). Since the histograms in (c) and (d) contribute to A$\to$B and \blue{B$\to$A} with equal weights, the symmetry also holds in (b).
    (e-h) As in (a-d) but for the driven data. The black line in (e) is now only a \textit{pseudo} potential. The symmetry of (g) and (h) does not imply symmetry in (f) since the weights of (g) and (h) are $96\ \%$ and $4\ \%$ for A$\to$B, and $12\ \%$ and $88\ \%$ for \blue{B$\to$A}, respectively.}
    \label{fig:transition times}
\end{figure}

The transition-path time symmetry has seemingly not yet been experimentally verified (although its violations in multidimensional non-equilibrium systems have already been observed experimentally \cite{Gladrow2019NC} and by computer simulations \cite{Ryabov2019JPCC}). Here we evaluate the transition times between two milestones (here points) $A$ and $B$ (see Fig.~\ref{fig:transition times}a,e) and evaluate frequency histograms of transition-path times as an estimator of their probability density for equilibrium (see Fig.~\ref{fig:transition times}a-d) and NESS dynamics (see Fig.~\ref
{fig:transition times}e-h). Note that since the original transition-path time symmetry does \emph{not} concern periodic systems, it only holds for transitions $A\to B$ and $B\to A$ \emph{either} passing through the middle (around $x=2\ \mu$m) \emph{or} passing through the periodic boundary ($x=0\ \mu$m), see overlap of blue and orange in Figs.~\ref{fig:transition times}c,d, respectively. Note that the symmetry is equally expected to hold in one-dimensional NESS \cite{Berezhkovskii2019JCP}, as confirmed in Fig.~\ref{fig:transition times}g,h (it only breaks down in \emph{multidimensional NESS}, see Ref.~\cite{Gladrow2019NC}).
Nevertheless, we see in Fig.~\ref{fig:transition times}b that the symmetry also holds for ``mixed'' transitions between $A$ and $B$, i.e., if we consider all transitions indifferent to whether they do or do not make use of the periodic boundary. This is generally the case for periodic equilibrium dynamics (given 1d Markov) since orange and blue in Fig.~\ref{fig:transition times}b are a weighted average of orange and blue in Figs.~\ref{fig:transition times}c,d where both colours are weighted with the same proportions [different weights cannot occur in equilibrium since this \blue{would lead to a contraction as it implies} directed motion ($\js\ne0$)].
The symmetry is violated for mixed transitions in Fig.~\ref{fig:transition times}f, since the weighting of Figs.~\ref{fig:transition times}g,h to obtain Fig.~\ref{fig:transition times}f is \emph{not} the same for orange and blue.

To summarize, Fig.~\ref{fig:transition times} demonstrates the validity of the predicted transition-path time symmetry between two milestones (Fig.~\ref{fig:transition times}c,d,g,h). While the transition-path time symmetry does not apply to periodic NESS dynamics (Fig.~\ref{fig:transition times}f), it holds true (in addition to the ``unmixed'' transitions in Fig.~\ref{fig:transition times}c,d,g,h) for periodic equilibrium dynamics (Fig.~\ref{fig:transition times}b).

\section{Conclusions}
We interrogated path-wise properties of driven colloids in a periodic
light field to experimentally demonstrate some highly intricate and insightful
features of transport, kinetics, and transition-path time statistics out of
equilibrium. Our main goal was to emphasize how these observables 
can be systematically utilized to critically assess 
and verify the quality of the experimental data.
\blue{We have reiterated that, with sufficient control and sampling, even subtle theoretical predictions can be rigorously tested and quantitatively validated. We hope this will motivate the experimental soft-matter community to undertake such tests and further enhance the quality of experimental data.}

Beyond the previous analyses of first-passage observables in
experimental data \cite{Zunke,Thorneywork2020SA} we compared the mean
particle velocity with the corresponding
first-passage result \cite{Reimann2001PRL} to check for short-time biases in 
non-equilibrium steady states. \red{Despite the measurement time being long enough for trajectories to equilibrate within one period $[0,L)$}, we
illustrate that observables, such as the one-sided mean first-passage
time in equilibrium, have \emph{not} yet converged, highlighting
the intricate features of first-passage times \blue{(and more generally, path-based observables).  That is, small statistical fluctuations (i.e., small error bars) do \emph{not} necessarily imply a sufficient quality of data as exemplified by the short-time biases in Fig.~\ref{fig:mFPT}. From a practical perspective, our results show that 
first-passage analysis can be used
to systematically test experimental data for short-time biases.}

Moreover, we verify and extend predicted symmetries for the more subtle transition-path times. By showing that the symmetry persists in effectively one-dimensional non-equilibrium systems, we underscore the usefulness of violations of the transition-path-time symmetry to infer simultaneously broken
time-reversal symmetry and the presence of multiple transition
pathways \cite{Berezhkovskii2019JCP,Gladrow2019NC}. 

\blue{We also highlighted two well-established but distinct, yet mathematically equivalent, formulations of the Langevin equation of motion for colloidal particles. Notably, the second formulation, while less commonly employed in experiments, provides a more thermodynamically expressive framework. We hope this will encourage further integration between soft-matter experiments and stochastic thermodynamics, particularly in advancing thermodynamic inference using speed limits and thermodynamic uncertainty relations.}

\section*{Author Contributions}
AG and SUE conceptualized the project, MAES and AG were responsible for the administration. SUE and AG provided the resources and acquired the funding for the experimental and theoretical work, respectively. CD and AG developed the theory. CD, YM, ABZB and MAES performed the investigations and validation. MAES and CD developed the software. ABZB and YM worked on the experimental setup. YM, CD and MAES worked on the methodology, data curation, and visualization. CD and AG wrote the original draft. CD, AG, ABZB, YM and MAES reviewed and edited the final version of the manuscript.

\section*{Conflicts of interest}
There are no conflicts to declare.

\section*{Acknowledgements}
Financial support from the European Research Council (ERC) under the European Union’s Horizon Europe research and innovation program (grant agreement No 101086182 to AG) 
\red{and from the Deutsche Forschungsgemeinschaft (DFG), project number 459399860 (to SUE and MAES),}
is gratefully acknowledged. This work is dedicated to the memory of our friend, mentor and colleague, Stefan U. Egelhaaf.
\\

\appendix
\blue{\section{Details on the periodic potential}
\label{sec:appendix}
For the above analysis, we determined period $L$ by comparing histograms for $x_\tau\equiv X_\tau\;{\rm mod}\;L$ over the left and right half, respectively, of particles in the field of view. The chosen value correct value of $L$ was optimized such that 
the two histograms optimally agree. While this does not rely on any assumptions about the exact shape of the potential (the shape was only later inferred from histograms; we only assumed that there is one barrier per period to remove ambiguity, since any $L$-periodic system is, e.g., also $2L$-periodic), this procedure may generally only serve as a fine-tuning for the value for $L$. Therefore, we here also present an alternative, more systematic, approach to infer $L$ using an approximation for the shape of $U(x)$.

\red{Moreover, while the theory and inference of $U(x)$ in the main part of the paper address general periodic potentials, the experimentally realized $U(x)$ is well approximated by a cosine. To elaborate on the discussion of the inference of $U(x)$, and to show the spatial variation that is \emph{not} investigated in the paper (since it is lost after projecting onto one period; as done, e.g., to arrive at Fig.~\ref{fig:inferred_U_v0}b), in this Appendix we also show a complementary approach, where we assume a cosine-shape and infer the amplitude and its spatial variation.

Following earlier experiments with comparable setups, we assume that the effective potential induced by the periodic light field can be well }approximated by \cite{Stefan4,Capellman2017}, $U(x)=U_0\cos(kx)+U_{bg}$,
with $k=2\pi/L$, $U_0$ the potential amplitude and $U_{bg}$ a constant background contribution. The potential background $U_{bg}$ becomes irrelevant as it is neglected by the derivative in Eq.~\eqref{equilibrium LE}. 
Over the whole field of view, particle density profiles $\rho(x)$ for the equilibrium trajectories were analyzed to extract the periodicity \red{and amplitude} of an underlying potential via $\rho(x)\propto\exp[-U(x)/k_{\rm B}T]$.
The natural logarithm of the density, $\ln (\rho(x))$, was evaluated, and its derivative, $\frac{d}{dx} \ln(\rho(x))$, to remove any background contributions. We used a sinusoidal function to fit the derivative for a window consisting of two periods and performed the same analysis for the entire dataset, covering 30 windows. Averaging the results across all windows 
we estimate the periodicity \red{and potential amplitude} of the light field. 
The \red{insets in Fig.~\ref{fig:appendix} show histograms} of the periodicity \red{and amplitude} values, with the frequency count plotted versus the extracted periodicity \red{and amplitude} for all measurements. The red curve represents a Gaussian fit to the histogram. This allows for a statistical summary with the mean and standard deviation extracted from the fit, \red{and in particular shows the spatial variation}. The error bars in the plot represent the standard deviation of the extracted periodicity
values within each window.

The results of the inference of $L$ in Fig.~\ref{fig:appendix}a, based on approximating $U$ as a cosine, are in excellent agreement with the value $L=4.135\ \mu$m chosen in the main part by matching histograms in different parts in the field of view without assuming the shape of $U(x)$. \red{Moreover, the amplitude inferred in Fig.~\ref{fig:appendix}b match closely with the barrier heights of $2.76$ $k_{\rm B}T$ and $2.65$ $k_{\rm B}T$ in Fig.~\ref{fig:inferred_U_v0}b, keeping in mind that the barrier (difference of maximum and minimum) in the cosine potential is $2V_0$.}
\begin{figure}[ht!]
\begin{center}
    \includegraphics[width=.7\linewidth]{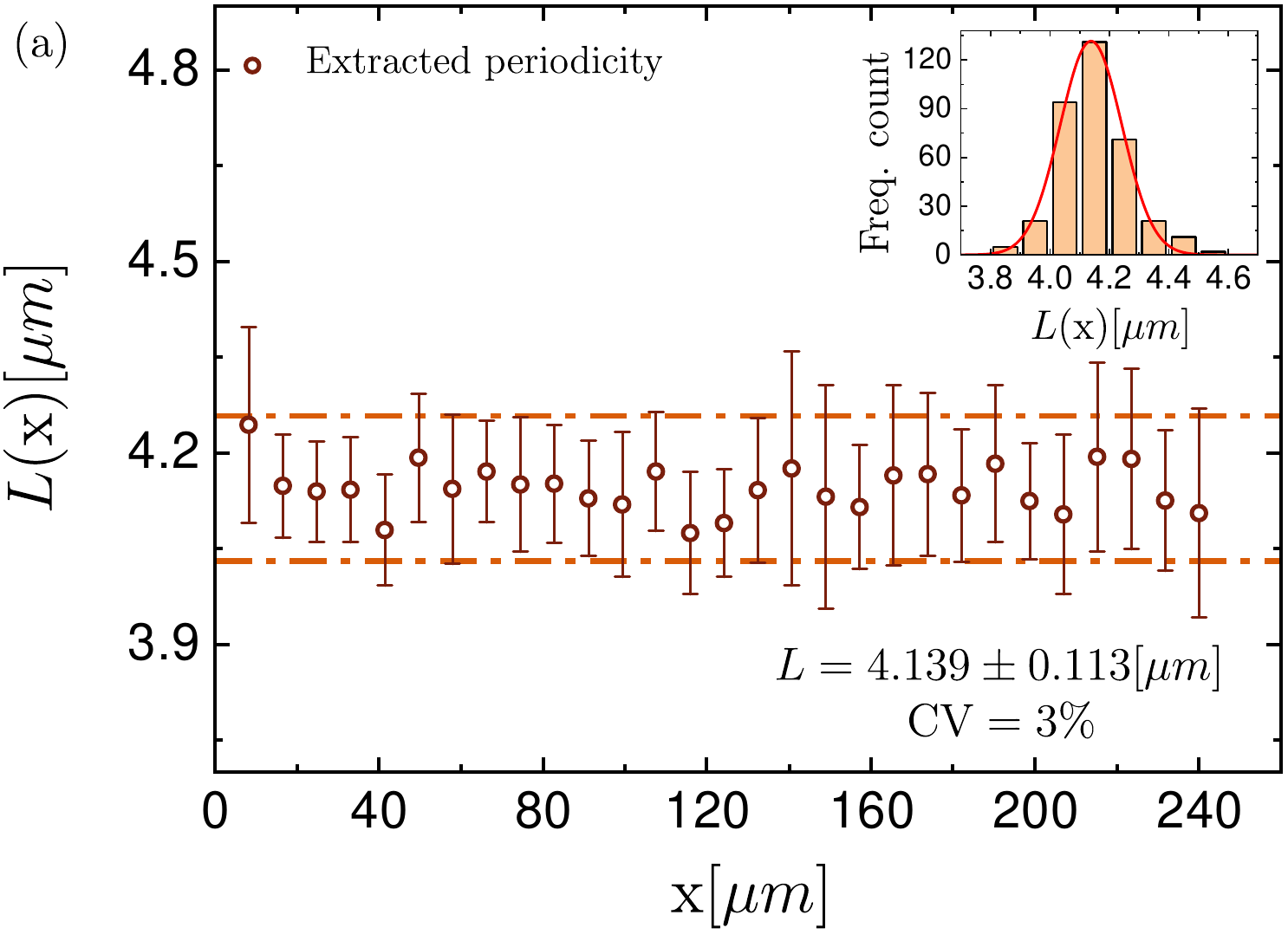}  \includegraphics[width=.7\linewidth]{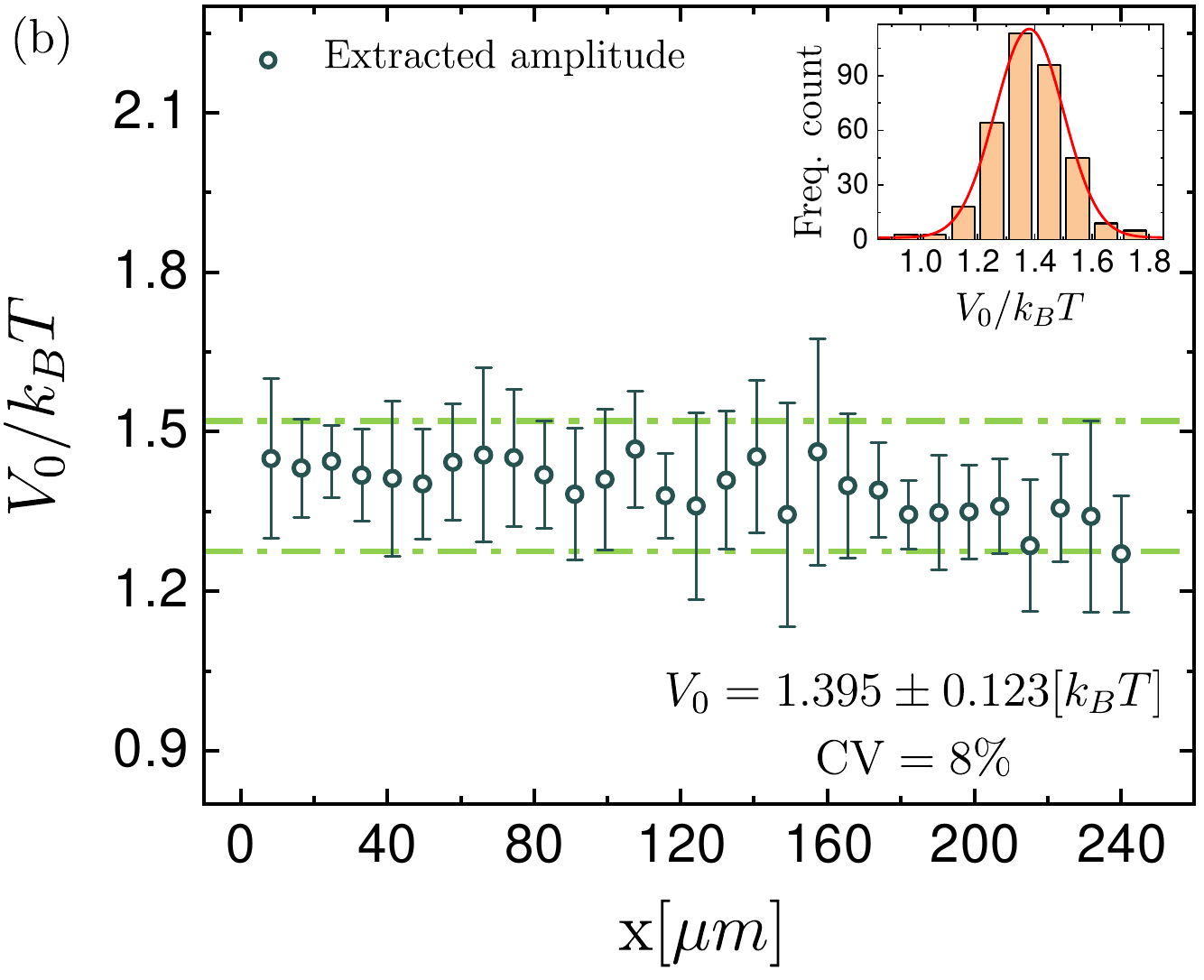}
\end{center}
\caption{\blue{
\red{(a)}
Extracted periodicity 
$L(x)$ as a function of position $x$ with error bars representing the standard deviation of $L(x)$ within each window. The dashed lines indicate the average value of the periodicity ($L=4.139 \mu m$) plus and minus one standard deviation ($\pm 0.113 \mu m$). The inset shows the histogram of the $L(x)$ values across all windows, with a Gaussian fit illustrating the distribution. The coefficient of variation (CV) is 3$\%$.
\red{(b) Extracted potential amplitude $V_0/k_{B}T$ with error bars showing the standard deviation within each window. The dashed lines represent the average amplitude $(V_0 = 1.395 \, k_{B} T)$ plus and minus one standard deviation$(\pm 0.123 \, k_{B}T)$. The inset displays the histogram of values across all windows, fitted with a Gaussian distribution. The coefficient of variation is 8$\%$.}
}}
\label{fig:appendix}
\end{figure}}

\balance


\bibliography{rsc1} 
\bibliographystyle{rsc} 

\end{document}